# Origin of exchange bias in [Co/Pt]$_{ML}$/Fe multilayer with orthogonal magnetic anisotropies

Sadhana Singh[1,2], Avinash G. Khanderao[1,3], Mukul Gupta[1], Ilya Sergeev[4], H. C. Wille[4], Kai Schlage[4], Marcus Herlitschke[4], Dileep Kumar[1,*]

[1]*UGC-DAE Consortium for Scientific Research, Khandwa Road, Indore,452017, India*
[2]*Nanomagnetism and Microscopy Laboratory, Department of Physics, Indian Institute of Technology Hyderabad, Kandi, Sangareddy, Telangana, 502285, India*
[3]*Department of Physics, Jagadamba Mahavidyalaya, Achalpur City, Maharashtra, 444806, India*
[4]*Deutsches Elektronen-Synchrotron DESY, Notkestraße 85, 22607 Hamburg, Germany*

*Corresponding Author Email: dkumar@csr.res.in; dileep.esrf@gmail.com

**Abstract.** Magnetization reversal of soft ferromagnetic Fe layer, coupled to [Co/Pt]$_{ML}$ multilayer [ML] with perpendicular magnetic anisotropy (PMA), has been studied in-situ with an aim to understand the origin of exchange bias (EB) in orthogonal magnetic anisotropic systems. The interface remanant state of the ML is modified by magnetic field annealing, and the effect of the same on the soft Fe layer is monitored using the in-situ magneto-optical Kerr effect (MOKE). A considerable shift in the Fe layer hysteresis loop from the centre and an unusual increase in the coercivity, similar to exchange bias phenomena, is attributed to the exchange coupling at the [Co/Pt]$_{ML}$ and Fe interface. The effect of the coupling on spin orientation at the interface is further explored precisely by performing an isotope selective grazing incident nuclear resonance scattering (GINRS) technique. Here, the interface selectivity is achieved by introducing a 2 nm thick Fe$^{57}$ marker between [Co/Pt]$_{ML}$ and Fe layers. Interface sensitivity is further enhanced by performing measurements under the x-ray standing wave conditions. The combined MOKE and GINRS analysis revealed the unidirectional pinning of the Fe layer due to the net in-plane magnetic spin at the interface caused by magnetic field annealing. Unidirectional exchange coupling or pinning at the interface, which may be due to the formation of asymmetrical closure domains, is found responsible for the origin of EB with an unusual increase in coercivity.

# I. Introduction

Exchange bias (EB), refers to the shift of hysteresis loop along the magnetic field axis when antiferromagnetic (AFM)/ ferromagnetic (FM) system is field cooled (FC) below the Nèèl's temperature ($T_N$) of the AFM layer [1,2,3]. This occurs due to unidirectional anisotropy induced in the FM layer owing to exchange interaction at the interface [1,2,3]. Hard/Soft FM bilayers exhibit a similar phenomenon due to the pinning of soft layer magnetization by the hard layer [4,5,6,7,8]. However, this pinning is a collective effect of direct exchange interaction at the interface and dipolar interaction due to stray fields [4,5,6,8]. Depending on the direction of field cooling [2,3,9,10,11,12] of AFM layer or saturation of hard layer [4,5,6,7,8,13,14,15], in-plane [2,4,5,6,7] or perpendicular EB [2,9,10,11,13,14,15] can be induced in magnetic thin films. In most studies, the AFM and hard FM layer have magnetic anisotropy co-planar to the magnetic anisotropy of the soft FM layer [2,4,5,6,7]. However, there are few studies which report occurrence of EB in system with orthogonal magnetic anisotropy (OMA) such as [Co/Pt]$_n$/NiFe [16,17,18,19], GaMnAsP/GaMnAs [20], Co$_{0.66}$Cr$_{0.22}$Pt$_{0.12}$ /Ni [21], Py/YIGBiLuAl [22], [Co(0.2)/Pd(1)]$_5$/CoFeB [12], [Pd/Co]/Cu/Co/Cu/[Co/Pd][23] etc. Here, the soft FM layers have in-plane magnetic anisotropy (IMA), whereas the hard layers have perpendicular magnetic anisotropy (PMA).

The appearance of EB in such systems was first reported by J. Sort *et al.* [16] in 2004. According to them, in-plane EB can be induced in [Co/Pt]$_n$/NiFe system having OMA by applying a very high in-plane magnetic field. Later on, similar results were reported in various other systems as listed above [12,17,18,19,21,22,23]. It was concluded that the orientation of spins at the interface between the soft IMA layer and hard PMA layer plays the main role in the appearance of EB. Application of large magnetic field changes the interface domain structure which in turn affects the orientation of spins at the interface, thereby leading to pinning of the soft layer and EB. As per A. Bollero *et al.* [18], asymmetry and stability of vortices between the upward and downward domains during the magnetization reversal might play an important role in inducing EB. It is reported that EB can be further tuned by modifying the strength of the PMA layer through the buffer layer [18], patterning long parallel strips instead of the continuous film [24] or applying an external perpendicular magnetic field [12]. Recently, a similar EB was reported in FM semiconductor GaMnAsP/GaMnAs bilayer with OMA [20] when FC in the strong in-plane magnetic field.

Although most of the above studies point out the importance of orientation of the interface spins for the origin of EB in OMA systems, a proper depth selective study to probe the interface magnetism is missing. So far, only indirect magnetic characterization techniques such as magneto-optical Kerr effect (MOKE), vibrating sample magnetometer (VSM) [16,18,19,21], Hall effect [17,20] and magnetic force microscopy (MFM) [16,17,18], as well as micromagnetic theoretical simulations [17,18,22] have been utilized to explain the origin of EB in these systems. However, these techniques provide average information on the magnetic behaviour of the multilayer and are not depth selective. Thus, they do not provide direct information on the orientation of moments at the interface. Since the interface plays a significant role, a detailed depth-resolved study is required to determine the net magnetization direction at the IMA and PMA layer interface to understand the origin of EB. Depth-sensitive techniques like conversion electron Mössbauer spectroscopy (CEMS), polarized neutron reflectivity (PNR) [21] and x-ray magnetic circular dichroism (XMCD) are more informative and powerful methods for interface-mediated studies. However, each has its limitations, such as limited penetration depth, low resonance counts, large sample size, long measurement time, etc. [25], thus, it cannot be helpful to probe a particular interface in a multilayer system. In such cases, synchrotron-based grazing incident nuclear resonance scattering (GINRS) with isotope sensitivity and high scattering yield can be used as an effective tool for depth-resolved measurement by placing a marker layer of the isotope at the designated depth.

Considering the above points, we have studied the origin of EB in [Co/Pt]$_{ML}$/Fe multilayer, where [Co/Pt]$_{ML}$ is prepared with strong PMA and Fe with in-plane anisotropy. In-situ MOKE is performed after annealing multilayer at different temperatures in the magnetic field. Magnetization reversal of Fe layer coupled to [Co/Pt]$_{ML}$ is studied using in-situ MOKE after annealing samples to varying temperatures in the magnetic field. The results are also compared with the samples annealed without the magnetic field. Furthermore, GI-NRS measurement, analogous to the Mössbauer spectroscopy, was performed to get interface resolved magnetism and relative orientation of the Fe magnetic moment ($\mu_{Fe}$) at the interface. For this purpose, a thin probe layer of isotopic (Fe$^{57}$) was deposited at the interface between the [Co/Pt]$_{ML}$ and Fe layer [26]. In addition, the x-ray standing wave (XSW) technique under planar waveguide conditions was utilized to enhance resonantly scattered counts from the Fe$^{57}$ isotope marker layer. The present study provides direct evidence of the spin structure at the interface and its role in inducing EB in the multilayer with two orthogonal anisotropy.

## II. EXPERIMENTAL:

Pt(250 Å) /[Co(6 Å) /Pt(30 Å)]$_{10}$/Co (5 Å) /Fe$^{57}$ (20 Å) /Fe(70 Å) /Pt(30 Å) multilayer (fig.1a) with Fe$^{57}$ marker layer at [Co/Pt]$_{ML}$ - Fe interface was deposited at room temperature (RT) using magnetron sputtering technique on Si (111) substrate at base pressure 5.7× 10$^{-7}$ Torr. The multilayer will be denoted as [Co/Pt]$_{ML}$/Fe for ease of reading. It may be noted that a thin Co layer of about 4 Å thickness was deposited on top of [Co/Pt]$_{ML}$ (before Fe$^{57}$ deposition) to enhance PMA and avoid interdiffusion between Pt and Fe$^{57}$ layers directly [9]. High-density layers (Pt) of thicknesses 24 nm and 3 nm are deposited respectively as a buffer and capping to enhance PMA [13,27], avoid surface oxidation, and generate XSW to perform interface resolved studies [25,28,29, 30]. In the present case, the top two Pt layers act as walls of the planar waveguide, whereas the Co/Fe$^{57}$/Fe structure acts as a guiding layer [25,28,29,30]. A schematic of the formation of XSW in the guiding layer is shown in figure 1. Two pieces of the same multilayer sample were annealed at various temperatures in ultra-high vacuum (UHV) conditions. One of the samples was annealed in the presence of a 1.5 kOe in-plane magnetic field (designated as "H-annealing"), whereas the other sample was annealed in the absence of a magnetic field in identical vacuum conditions. MOKE in polar geometry (P-MOKE) was performed to confirm PMA in [Co/Pt]$_{ML}$ multilayer, whereas magnetization reversal of Fe layer is studied by collecting hysteresis loop after different annealing stages using longitudinal MOKE (L-MOKE) measurements. It may be noted that hysteresis loops were measured at RT in ± 250 Oe magnetic field, which is much less than the magnetic field required to disturb the magnetization reversal of [Co/Pt]$_{ML}$ multilayer due to PMA. Therefore, hysteresis loops presented in this work correspond to the soft magnetic structure on [Co/Pt]$_{ML}$ multilayer unless specified.

GINRS experiments were carried out under XSW using x-ray synchrotron radiation source at the P01, Dynamics Beamline at PETRA III, DESY, Hamburg, Germany [31]. Interface selectivity was achieved due to crossing XSW antinodes with the interface at an appropriate x-rays incident angle [30,32,33]. The correct angle of the incident (q= 0.066 Å$^{-1}$) to perform GINRS measurements is extracted based on electronic and nuclear reflectivity measurements. GINRS experiments are conducted in time mode with a beam having 40 bunch with a bunch separation of 192 ns. The x-ray energy was tuned to 14.41 KeV, the nuclear transition energy corresponding to the Fe$^{57}$ Mössbauer isotope with a natural lifetime of 141 ns.

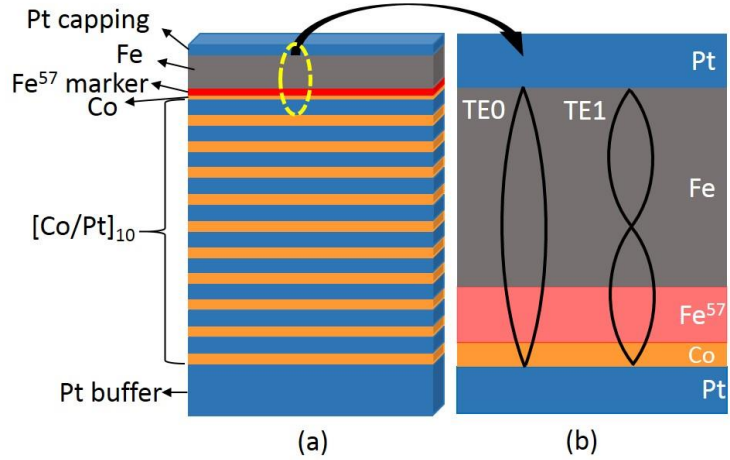

**Fig. 1:** Schematic diagrams showing (a) [Co/Pt]$_{ML}$/Fe multilayer structure and (b) enlarged view of XSW formation through planar waveguide structure between top two high-density Pt layers

# III. RESULT AND DISCUSSION

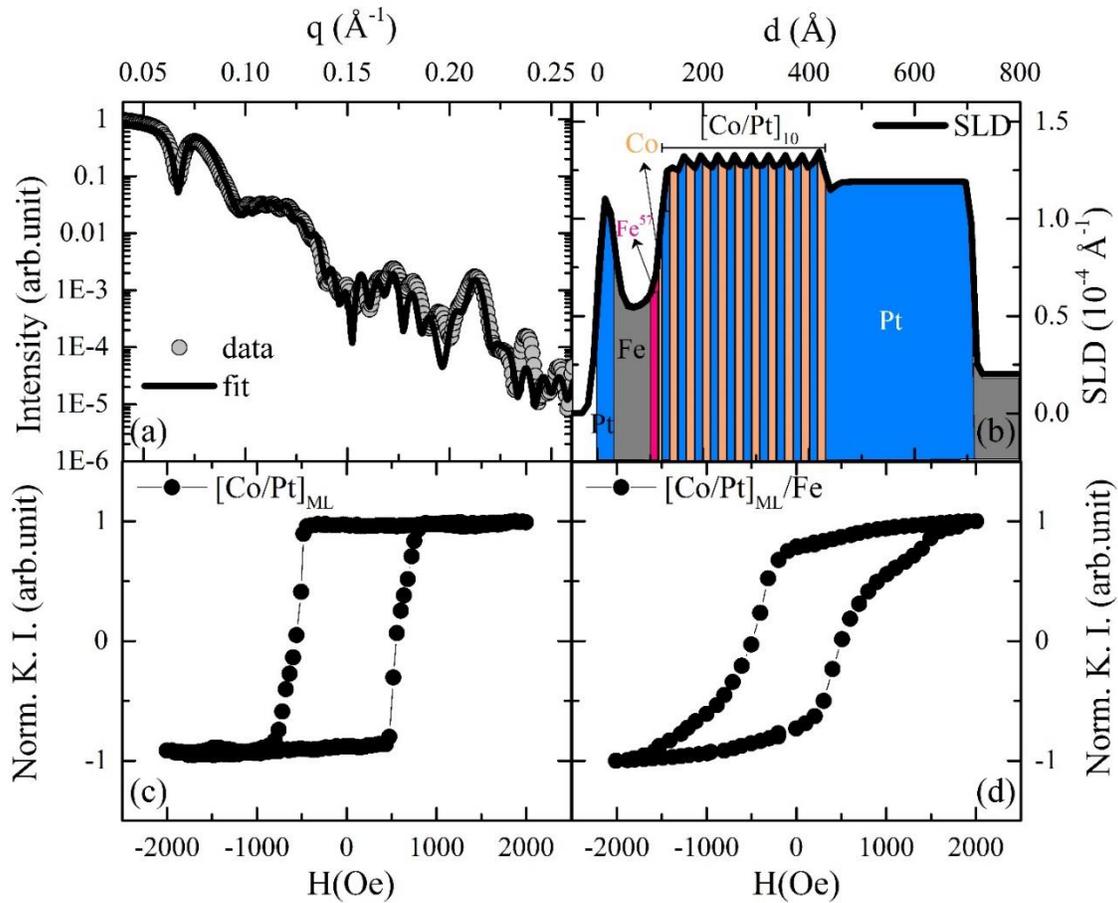

**Fig. 2:** (a) Fitted XRR pattern and (b) extracted scattering length density (SLD) profile of as deposited [Co/Pt]$_{ML}$/Fe multilayer. P-MOKE hysteresis loop of (c) [Co/Pt]$_{ML}$ and (d) [Co/Pt]$_{ML}$/Fe.

Figure 2(a) show the x-ray reflectivity (XRR) pattern (circles) of the as-deposited [Co/Pt]$_{ML}$/Fe multilayer fitted (solid line) using Parratt's formalism [34]. The corresponding extracted scattering length density (SLD) profile from XRR fitting is shown in fig. 2(b). Small oscillations in the XRR pattern signify the total thickness of the multilayer, whereas Bragg's peak at ~q (momentum transfer vector perpendicular to the surface) = 0.21 Å$^{-1}$ corresponds to the bilayer thickness of the [Co/Pt]$_{ML}$. After fitting, the thickness of Co and Pt layers in [Co/Pt]$_{ML}$ is found to be 4.6 Å (± 0.2Å) and ~26.5 Å (± 0.5 Å), respectively, whereas Fe layer thickness is about ~85 Å (including Fe$^{57}$ thickness). It may be noted that Fe$^{57}$ and Fe$^{Nat}$ are chemically the same (same electron density in both isotopes); therefore, the XRR technique gives combined information of Fe$^{57}$ and Fe$^{Nat}$ layers. The sharp dip around q = 0.066 Å$^{-1}$, below the critical angle of Pt, indicates the resonance coupling of incident x-rays resulting from forming the waveguide between two high-density Pt layers in top **Pt**/Co/Fe$^{57}$/Fe/**Pt** layers [25,28,29,30]. Final sample structure obtained after XRR fitting is Pt (240 Å) / [Co (5 Å) /Pt (26 Å)]$_{10}$/Co (4 Å) /Fe$^{57}$ (15 Å) /Fe (70 Å) /Pt (30 Å).

Figure 2(c) and 2(d) shows P-MOKE hysteresis loop of [Co/Pt]$_{ML}$ and [Co/Pt]$_{ML}$/Fe multilayer, separately. It is observed that [Co/Pt]$_{ML}$ multilayer exhibits an almost square hysteresis loop with very high remanence (Mr=0.94) and low coercivity (H$_C$ ~560 Oe). It clearly shows that the [Co/Pt]$_{ML}$ multilayer exhibits strong PMA. The origin of PMA in this multilayer is known to originate from the spin-orbit coupling between Co and Pt atoms [35,36]. The hysteresis loop of [Co/Pt]$_{ML}$/Fe is expected to be a combination of two loops. A square loop corresponds to the [Co/Pt]$_{ML}$ with a magnetic easy axis along the direction of the applied field, and the slanted loop corresponds to the Fe layer with an easy axis along the in-plane direction. Thus, the magnetic reversal of [Co/Pt]$_{ML}$ occurs in relatively less field than the Fe layer. In the present case hysteresis loop of [Co/Pt]$_{ML}$/Fe needs a larger field (>2000 Oe ) to achieve saturated magnetization reversal in this geometry. Because of the fact, with the magnetic field strength of ±2000 Oe, both the Fe layer and [Co/Pt]$_{ML}$ participate in the magnetic reversal process resulting in a combined loop in P-MOKE. Furthermore, the hysteresis loop of the [Co/Pt]$_{ML}$/Fe multilayer was measured in ±250 Oe at RT in L-MOKE geometry. In L-MOKE geometry, the magnetic field ±250 Oe is sufficient for the magnetization reversible of the Fe layer but is much less than the field required to switch [Co/Pt]$_{ML}$ structure.

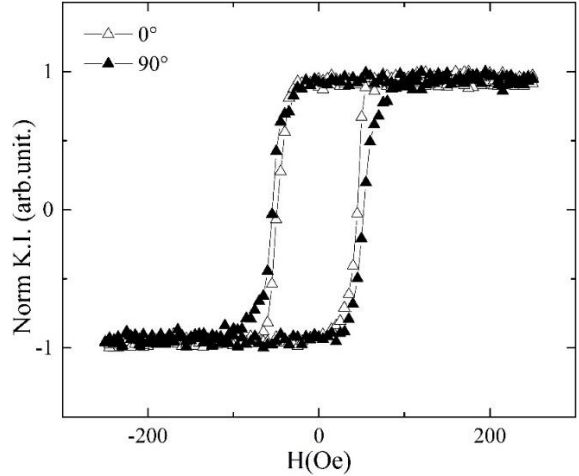

**Fig. 3:** L-MOKE hysteresis loop of [Co/Pt]$_{ML}$/Fe in ±250 Oe magnetic field along azimuthal angles θ= 0° and θ= 90° (θ= 0° is any arbitrary azimuthal direction).

As shown in figure 3, the loops exhibit an almost single-step square hysteresis loop along two in-plane azimuthal directions θ= 0° & 90°. Both loops are almost similar in shape, confirming isotropic Fe magnetism in the film plane. The loops are also centred in the magnetic field axis (x-axis), indicating the absence of preferential pinning in the film. Hence, no exchange bias (EB) is present in the as-deposited stage.

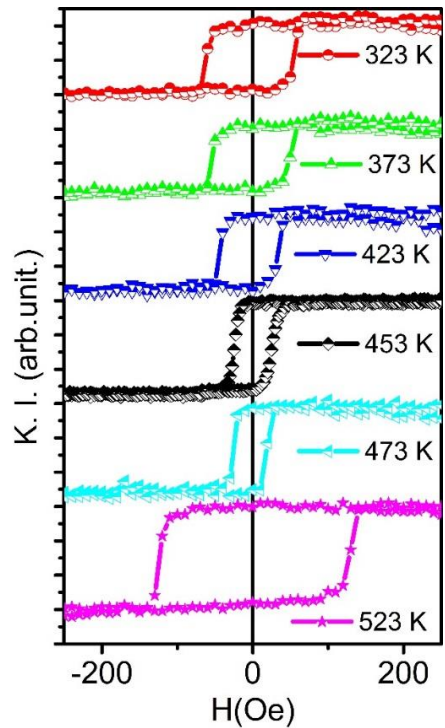

**Fig. 4:** L- MOKE hysteresis loop of [Co/Pt]$_{ML}$/Fe multilayer after annealing at different temperatures under UHV conditions.

Prior to the EB study, the thermal stability of the multilayer was investigated by collecting hysteresis loops of the multilayer layer in a magnetic field ±250 Oe (L-MOKE geometry) and XRR patterns after annealing sample at different temperatures up to 673K. All hysteresis loops are presented in figure 4. It is observed that with increasing annealing temperature, $H_C$ of multilayer first decreases up to 473 K. However, an unusual increase in $H_C$ was observed after annealing at 523 K. All the hysteresis loops remain centred along the field axis; Hence no preferential coupling at the interface of $[Co/Pt]_{ML}$ is induced after annealing. Initial decreases in $H_C$ with increasing annealing temperatures can be understood in terms of the removal of stress or defects which might have been generated during film growth. The unusual increase in $H_C$ after annealing at 523 K may be due to (i) increased interface domain wall pinning at the interfaces due to interdiffusion or (ii) an increase in isotropic coupling between hard $[Co/Pt]_{ML}$ and soft Fe layer.

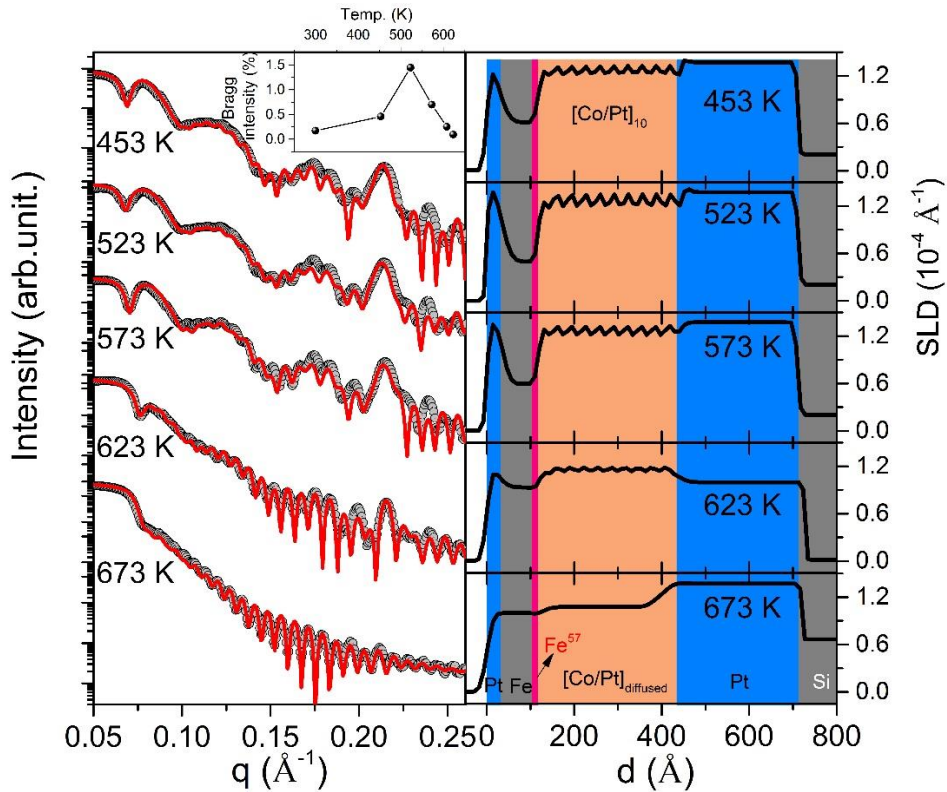

**Fig. 5:** (a) Fitted XRR pattern and (b) SLD profile of $[Co/Pt]_{ML}$/Fe multilayer after annealing it at different temperatures under UHV conditions. Inset gives variation in normalized Bragg peak intensity after annealing sample up to 673 K.

The XRR measurements of the multilayer were carried out after annealing at different temperatures ranging from 453 K to 673 K (see fig 5a) to correlate the structural properties (layer thickness, interface roughness, and interface mixing) with the evolution of magnetic properties after annealing. All XRR patterns are shown in figure 5(a), where solid lines represent the best fit to the data using Parratt's formalism [34]. The corresponding SLD profiles, extracted after fitting, are shown in fig 5b. The normalized Bragg peaks intensity increases with annealing upto 523 K (inset fig. 5a). With further annealing, it decreases at 623 K and finally disappears after annealing at 673 K. After annealing at this temperature XRR pattern consist of only the total thickness oscillations (Kiessig oscillations). As Bragg peak intensity is directly related to the interface roughness of the [Co/Pt]$_{ML}$[37,38], therefore the initial increase in Bragg peak intensity clearly suggests that the Co/Pt interfaces became sharp due to interface demixing up to 523 K. Sharpening of the [Co/Pt]$_{ML}$ interfaces can also be seen in SLD profiles corresponding to the temperatures 453K and 523 K. Decrease in Bragg intensity beyond 523 K is mainly due to diffusion at the interface across the layers of the multilayer. Complete intermixing of Co and Pt layers at 623 K results in the disappearance of the Bragg peak.

Similar observations have also been reported in the literature where such multilayers are stable up to ~ 673 K. Interdiffusion occurs with further increase in the temperature [39,40,41,42,43], which is responsible for decreased PMA in Co/Pt multilayers system. Thus, from XRR, it is clear that the increase in H$_C$ at 523 K does not originate due to the interface domain pinning. However, a decrease in the interface roughness may increase the PMA of [Co/Pt]$_{ML.}$ It, in turn, increases the strength of the coupling between hard [Co/Pt]$_{ML}$ and soft Fe layer resulting in enhanced H$_C$ at 523 K.

Figure 6(a) shows hysteresis loops of the sample measured at RT after H-annealing at different temperatures in the presence of ~1.5 kOe magnetic field. On H-annealing up to 453 K, hysteresis loop remains centred at the magnetic field axis. However, for temperatures 523 K and above, hysteresis loops are shifted from their origin. This shift is higher with a further increase in temperature up to 573 K. This observation clearly suggests the appearance of the EB at these H-annealed states. The figure 6(b-c) shows the variation of HC and HEB with increasing temperature for qualitative understanding. H$_{EB}$ Vs temperature in fig. 6(c) gives the slow increase of EB up to 450K. But with a further increase in H-annealing temperature up to 573K, it drastically increases from H$_{EB}$=6Oe to 50 Oe. In fig 6(b), Hc variation shows a drastic increase in the Hc beyond H-annealing at ~450K. Its value is almost double compared to the annealing in the absence of field. It is clear that modified interface spin structure, due to the H-annealing, is the main factor that plays an important role in the origin of EB in this sample.

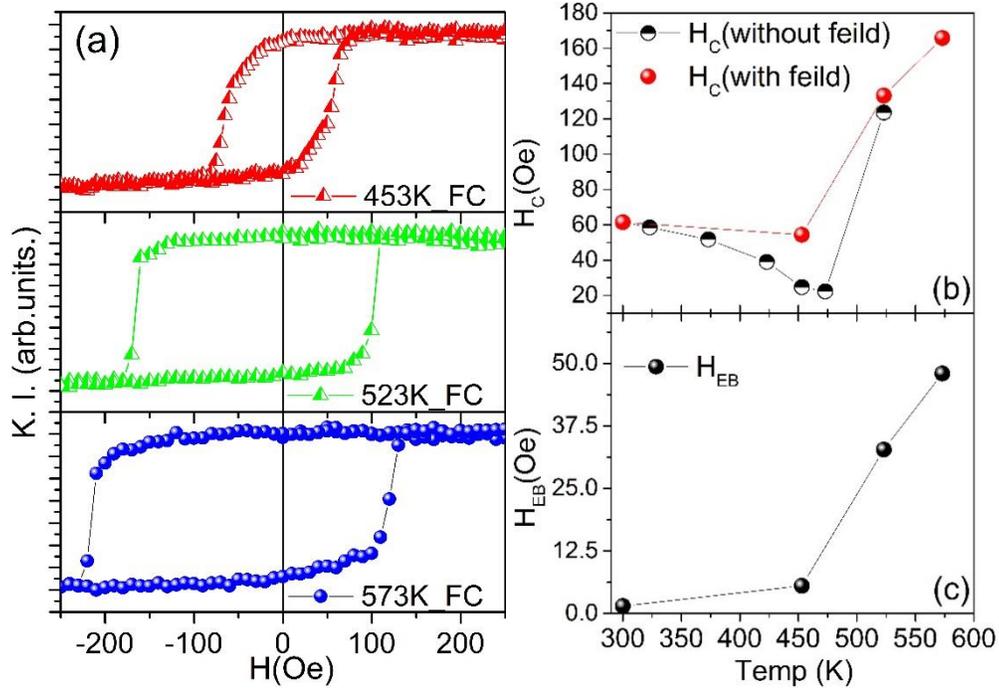

**Fig. 6:** (a) Hysteresis loop of [Co/Pt]$_{ML}$/Fe multilayer after annealing in the presence of ~ 1500 Oe magnetic field. Variation of (b) H$_C$ (with and without field annealing) and (c) H$_{EB}$ with increasing annealing temperature.

To investigate role of interface magnetism, depth-selective GI-NRS measurements are performed to probe magnetism (hyperfine field and spin orientation) at the interface for H-annealed samples. As GI-NRS is an isotope-sensitive technique, a thin Fe$^{57}$ layer is used as a marker layer at the [Co/Pt]$_{ML}$ and Fe interface. Since the Fe$^{57}$ layer is very thin, the XSW technique [25,28,29,30,44] is utilized to enhance the resonance counts from the marker layer at the interface. In the XSW technique, x-ray electric field intensity (EFI) is confined in different XSW modes (nodes and antinodes). By varying the angle of incidence, the position of the antinodes can be moved along the z-direction (depth) [30,32,33]. Enhanced nuclear resonance yield can be achieved by moving the antinode position across the marker layer at a particular incident angle. Based on the final sample structure, as obtained from the XRR measurements, x-ray EFI inside the guiding layer was calculated as a function of scattering vector q (incident angle θ) using Parratt's formalism [34] and shown in fig. 7(a).

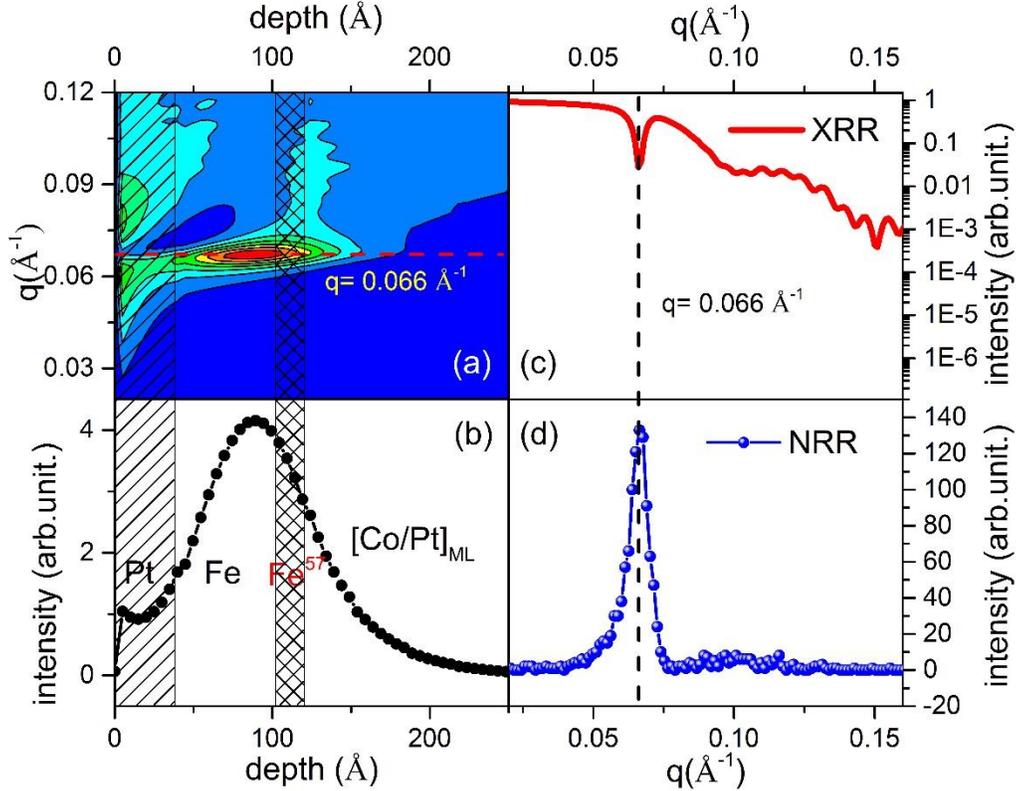

**Fig. 7:** (a) Contour plot of x-ray field intensity inside $[Co/Pt]_{ML}/Fe$ multilayer as a function of depth and scattering vector q. (b) Electric field intensity (EFI) extracted for an angle of incidence of q=0.066 Å$^{-1}$. XSW antinode (TE0 mode) crosses the Fe57 marker layer at this angle. Experimental (c) non-resonant and (d) resonant reflectivity of the multilayer showing minima and maxima respectively at q=0.066 Å$^{-1}$.

A clear confinement of the EFI (TE0 mode) was observed at around the angle of incidence θ=0.259° (q=0.066Å$^{-1}$) and crosses the Fe$^{57}$ layer at the interface. It suggests that GINRS measurement at q= 0.066 Å$^{-1}$ will increase the nuclear resonance counts from the Fe$^{57}$ layer. To further confirm it experimentally, nuclear resonance reflected intensity is collected as a function of q along with XRR and shown in fig. 7(c) and 7(d). More than an order of higher resonant intensity has been observed at around q=0.066 Å$^{-1}$. In view of the above facts, all GI-NRS measurements (time spectra) were recorded at an incidence angle of q= 0.066 Å$^{-1}$ for H-annealed multilayer at various temperatures. For the measurement, the sample was placed in the path of the beam such that the direction of the magnetic field applied during annealing is normal to the scattering plane of the incident beam. Fig. 8(a) shows a schematic representation of sample orientation with respect to the x-ray beam. Figure 8(b) gives measured GINRS time spectra. To get magnetic information, the measured time spectra were fitted using the simulation and least-squares fitting procedure of the REFTIM program [45,46] by taking parameters as obtained from

XRR and NRR fitting. The best fit to the time spectra was obtained by dividing the Fe$^{57}$ layer into two layers with three overlapping hyperfine fields (Bhf) components, i.e. 32.9 ±0.02 T, 32.6 T ±0.015 T, 28.7±0.79 T.

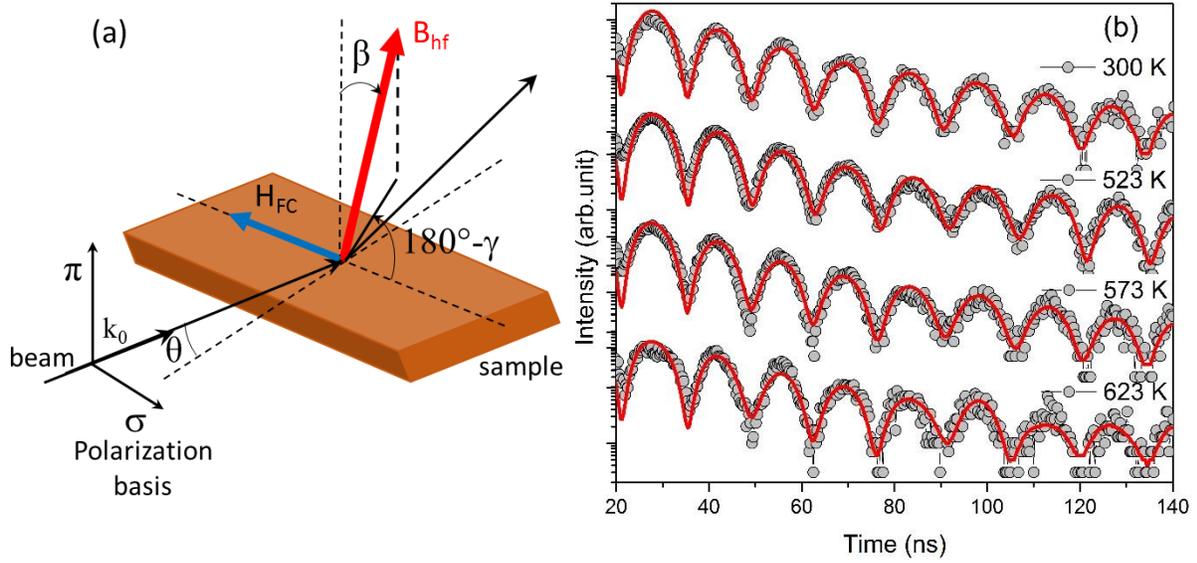

**Fig. 8:** (a) Schematic representation of the orientation of sample with respect to the scattering plane and magnetic field applied during annealing. θ is the angle of incidence of the beam with respect to the sample surface. (b) GI-NRS time spectra of [Co/Pt]$_{ML}$/Fe multilayer after annealing at various temperatures and cooling to RT in the presence of an in-plane magnetic field (~1500 Oe ). Spectra are fitted using REFTIM software [45,46].

**Table I:** Fitting parameter of GI-NRS spectra obtained using REFTIM software [45,46].

| Temp. | Bhf(T) | β(°) | γ(°) |
|---|---|---|---|
| 300 K | 32.89 | 2.0 | 2.0 |
|  | 32.61 | 0.0 | 2.0 |
|  | 28.75 | 0.0 | 2.0 |
| 523 K | 32.85 | 14.0 | 2.0 |
|  | 32.75 | 11.0 | 2.0 |
|  | 28.25 | 8.6 | 2.0 |
| 573 K | 32.90 | 16.0 | 2.0 |
|  | 32.61 | 11.0 | 2.0 |
|  | 28.29 | 13.0 | 2.0 |
| 623 K | 32.94 | 34.4 | 2.0 |
|  | 32.44 | 34.3 | 2.0 |
|  | 27.78 | 34.3 | 2.0 |

It is found that the interface layer has a higher contribution (~27%) of the component with reduced hyperfine field (28.7 ±0.79 T) due to interfacial mixing as compared to the upper interface layer (6%). The fitting parameters are tabulated in Table I. It was found that at RT (300 K), spins of the $Fe^{57}$ layer are oriented perpendicular to the film plane. On annealing the multilayer at 523 K in the presence of a magnetic field, it was observed that net magnetization rotates by approximately 11° with respect to the plane normal (β) along the saturation field direction (γ). The net magnetization further rotates in the film plane with increasing temperature. The rotation of moments in the in-plane direction may be attributed to magnetic field annealing.

As per literature [47,48,49], when a soft FM layer is deposited on a layer with strong PMA, for thicknesses below the exchange length, the soft FM layer moments are aligned perpendicular. With increasing thickness, the moments gradually rotate in-plane. Furthermore, a gradual rotation of moments occurs within the layer even for a constant thickness of the soft FM layer [49]. In the present case, $[Co/Pt]_{ML}$ has strong PMA; hence, moments of Fe at the interface will be aligned normal to the film plane due to dipolar coupling as observed through GINRS measurement for RT spectra. Thus, the absence of preferential pinning in the in-plane direction during the magnetization reversal of the IPA layer results in the absence of EB in the as-deposited sample. This also explains the absence of EB when the multilayer is annealed without field. A strong in-plane field is required to rotate these moments in-plane to observe EB [12,16,17,21,22]. However, in the present study, magnetization reversal was studied as a function of H-annealing in 1500 Oe magnetic field. Application of in-plane magnetic field during annealing might create magnetic spins at the interface saturated along the direction of the applied magnetic field [50,51,52,53].

However, since the field is small compared to that required to reverse the $[Co/Pt]_{ML}$, a complete reversal of its moments in the in-plane direction will not occur [50]. This creates a net remanent in-plane magnetization resulting in unidirectional anisotropy at the interface that pins the IPA layer and results in the appearance of EB (see fig. 9). This condition is equivalent to AFM/FM bilayer where uncompensated moment at the interface pins the FM layer. GINRS study confirms the presence of magnetic spins at the interface with net in-plane remanent magnetization along the direction of the applied magnetic field. Rotation of moments in the direction of the applied field is in accordance with the literature, wherein H-annealing in the presence of a moderate magnetic field at higher temperature assists the ordering of spin in the direction of the applied field [50,51,52,53]. As shown in fig. 9a, the remanent in-plane magnetization may result from asymmetrical closure domains at the interface parallel to the applied field [16,17,18,19,20]. This condition is otherwise absent when moments at the interface have a perpendicular orientation or when the closure domains are oriented uniformly (as shown in fig. 9b). Below 523K, thermal energy coupled with 1500 Oe magnetic field may not be sufficient to orient the spins in-plane, resulting in the absence of EB. Therefore EB is not present in these states of the sample. The drastic increase in the coercivity is explained by the increase in effective pinning of the Fe layer by the PMA layer.

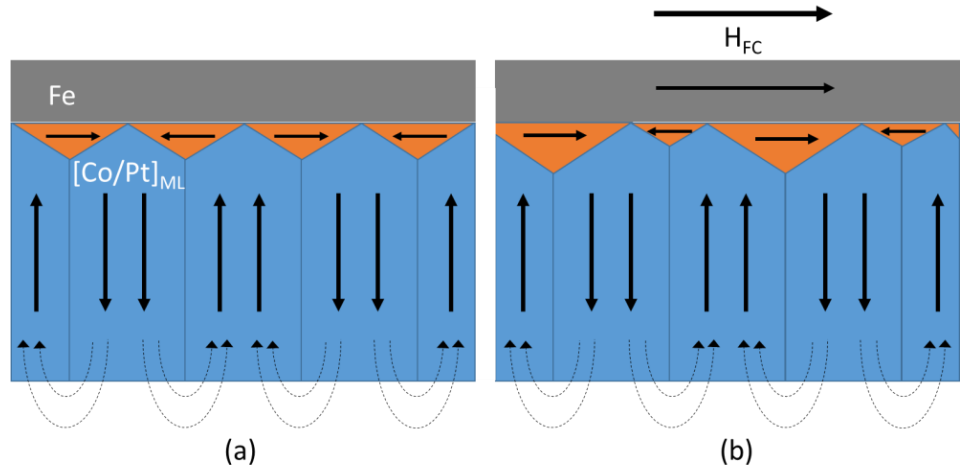

**Fig. 9:** Schematic representation of closure domains for the sample annealed (a) in the absence of magnetic field, and (b) in the presence of 1500 Oe magnetic field in the plane of the film.

Thus, the observed unusual increase in $H_C$ and appearance of EB at 523 K in [Co/Pt]ML/Fe with orthogonal anisotropy is attributed to two factors. (i) Decrease in interface roughness which leads to the increase in the strength of PMA in Co/Pt multilayer, and (ii) H-annealing induced net in-plane remanent magnetization at the interface that results in unidirectional anisotropy.

## IV. Conclusion:

Magnetization reversal of soft ferromagnetic Fe layer, coupled to [Co/Pt]$_{ML}$ multilayer with perpendicular magnetic anisotropy (PMA), has been studied in-situ with an aim to understand the origin of exchange bias (EB) in such orthogonal magnetic anisotropic (OMA) systems. Along with in-situ MOKE and XRR studies, interface resolved magnetic properties are investigated at synchrotron radiation sources using isotope selective GI-NRS technique. A thin Fe$^{57}$ isotope was deposited at the interface to study interface magnetism, whereas interface to bulk contribution is further increased by performing GINRS measurements under x-ray standing wave conditions.

Magnetic orientation and its magnitude at the interface of the two orthogonal magnetic anisotropic layers are studied after modifying the interface remanant state by annealing in the magnetic field. The combined MOKE and GINRS analysis revealed that [Co/Pt]$_{ML}$/Fe multilayer exhibits an unusual increase in $H_C$ and appearance of EB when annealed in the presence of the in-plane magnetic field; otherwise, it is found absent. This unusual behaviour is attributed to (i) a decrease in interface roughness which leads to the increase in the strength of PMA in Co/Pt multilayer, and (ii) magnetic field induced net in-plane remanent magnetization at the interface that results in unidirectional anisotropy the EB in the multilayer due to the net in-plane magnetic spin at the interface caused by field annealing. Unidirectional exchange coupling or pinning at the interface, which may be due to the formation of asymmetrical closure domains, is found responsible for the origin of EB with an unusual increase in coercivity.

## Acknowledgement:

We acknowledge Dr. V. Raghavendra Reddy and Er. Anil Kumar, UGC-DAE CSR, Indore, India, for polar MOKE and XRR measurements. We acknowledge Deutsches Elektronen-Synchrotron (DESY) (Hamburg, Germany), a member of the Helmholtz Association HGF, for providing experimental facilities. I would also like to thank the Department of Science and Technology (DST), Government of India, for providing financial assistance to perform experiments at DESY, Hamburg, within the framework of the India@DESY collaboration (Proposal No. I-20160350). We acknowledge Er. Layant Behera and Prabhat Kumar, UGC-DAE CSR, Indore, India, for film deposition using sputtering technique.

## Reference:

[1] W.H. Meiklejohn, C.P. Bean, Phys. Rev. **102**, 1413 (1956); https://doi.org/10.1103/PhysRev.102.1413

[2] J. Nogues and I. K. Schuller, J. Magn. Magn. Mater. **192**, 203 (1999); https://doi.org/10.1016/S0304-8853(98)00266-2

[3] D. Kumar, S. Singh, and A. Gupta, J. Appl. Phys. 120, 085307(2016); https://doi.org/10.1063/1.4961521

[4] A. Berger, D. T. Margulies, and H. Do, Appl. Phys. Lett. **85**, 1571 (2004); http://dx.doi.org/10.1063/1.1787161

[5] A. Berger, D. Margulies, H. Do, A. Ktena, and K. Dahmen, J. Appl. Phys. **97**, 10K109 (2005); http://dx.doi.org/10.1063/1.1854274

[6] Z. Hussain, D. Kumar, V.R. Reddy, A. Gupta, J. Magn. Magn. Mater. **430**, 78–84 (2017); https://doi.org/10.1016/j.jmmm.2017.01.052

[7] S. Polisetty, S. Sahoo, A. Berger, and C. Binek, Phys. Rev. B **78**, 184426 (2008); https://doi.org/10.1103/PhysRevB.78.184426

[8] S. Singh, D. Kumar, B. Bhagat, R. J. Choudhary, V.R. Reddy, J. Phys. D: Appl. Phys., **51**, 075006 (2018); https://doi.org/10.1088/1361-6463/aaa539

[9] J. Sort, V. Baltz, F. Garcia, B. Rodmacq, and B. Dieny, Phys. Rev. B 71, 054411 (2005); https://doi.org/10.1103/PhysRevB.71.054411

[10] S. Maat, K. Takano, S. S. P. Parkin, and Eric E. Fullerton, Phys. Rev. Lett. 87, 087202 (2001); https://doi.org/10.1103/PhysRevLett.87.087202

[11] S. M. Zhou, L. Sun, P. C. Searson, and C. L. Chien, Phys. Rev. B 69, 024408 (2004); https://doi.org/10.1103/PhysRevB.69.024408

[12] J. Jiang, T. Yu, R. Pan, Q.-T. Zhang, P.Liu, H. Naganuma, M. Oogane, Y. Ando and X. Han, Appl. Phys. Express **9**, 063003 (2016); https://doi.org/10.7567/APEX.9.063003

[13] S. T. Lim, M. Tran, J. W. Chenchen, J. F. Ying and G. Han, J. Appl. Phys. **117**, 17A731 (2015); https://doi.org/10.1063/1.4916295

[14] A. Baruth and S Adenwalla, J. Phys.: Condens. Matter **23**(37), 376002, (2011); https://doi.org/10.1088/0953-8984/23/37/376002

[15] A. Hierro-Rodriguez, a), J. M. Teixeira, M. Vélez, L. M. Alvarez-Prado, J. I. Martín, and J. M. Alameda, Appl. Phys. Lett. **105**, 102412 (2014); https://doi.org/10.1063/1.4895771

[16] J. Sort, A. Popa, B. Rodmacq, and B. Dieny, Phys. Rev.B. **70**, 174431 (2004); https://doi.org/10.1103/PhysRevB.70.174431

[17] A. Bollero, L. D. Buda-Prejbeanu, V. Baltz, J. Sort, B. Rodmacq, and B. Dieny Phys. Rev. B **73**, 144407 (2006); https://doi.org/10.1103/PhysRevB.73.144407

[18] A. Bollero, V. Baltz, L. D. Buda-Prejbeanu, B. Rodmacq, and B. Dieny, Phys. Rev. B **84**, 094423 ( 2011); https://doi.org/10.1103/PhysRevB.84.094423

[19] A. Bollero, L.D. Buda-Prejbeanu, V. Baltz ; B. Rodmacq ; B. Dieny, IEEE Trans. Magn. **42**, 2990 (2006); https://doi.org/10.1109/TMAG.2006.879758

[20] S. Choi, S.-K. Bac, X. Liu, S. Lee, S. Dong, M. Dobrowolska and J. K. Furdyna, Sci Rep **9,** 13061 (2019); https://doi.org/10.1038/s41598-019-49492-4

[21] D. Navas, J. Torrejon, F. Beron, C. Redondo, F. Batallan, B.P. Toperverg, A. Devishvili, B. Sierra, F. Castano, K.R. Pirota and C.A. Ross, New J. Phys. **14**, 113001 (2012); https://doi.org/10.1088/1367-2630/14/11/113001

[22] N. Vukadinovic, J. Ben Youssef, V. Castel and M. Labrune, Phys. Rev. B **79**, 184405 (2009); https://doi.org/10.1103/PhysRevB.79.184405

[23] P. Y. Yang, X. Y. Zhu, F. Zeng, and F. Pan, Appl. Phys. Lett. **95**, 172512 (2009); https://doi.org/10.1063/1.3257696

[24] A. Bollero, B. Dieny, J. Sort, K. S. Buchanan, S. Landis and J. Nogues, Appl. Phys. Lett. **92**, 022508 (2008); https://doi.org/10.1063/1.2833124

[25] A. G. Khanderao, I. Sergueev, H. C. Wille, and D. Kumar, Appl. Phys. Lett. **116**, 101603 (2020); https://doi.org/10.1063/1.5135361

[26] L. Bocklage, C. Swoboda, K. Schlage, H.-C. Wille, L. Dzemiantsova, S. Bajt, G. Meier, and R. Röhlsberger, Phys. Rev. Lett. **114**, 147601 (2015); doi: https://doi.org/10.1103/PhysRevLett.114.147601

[27] P. Chowdhury, P. D. Kulkarni, M. Krishnan, Harish C. Barshilia, A. Sagdeo, S. K. Rai, G. S. Lodha, and D. V. Sridhara Rao, J. Appl. Phys. **112**, 023912, (2012); doi: https://doi.org/10.1063/1.4739284

[28] A. Gupta, P. Rajput, A. Saraiya, V. R. Reddy, M. Gupta, S. Bernstorff, and H. Amenitsch, Phys. Rev. B **72,** 075436 (2005); https://doi.org/10.1103/PhysRevB.72.075436

[29] A Gupta, D Kumar, C Meneghini, J Zegenhagen, J. Appl. Phys. **101** (9), 09D117 (2007); https://doi.org/10.1063/1.2711698

[30] M. S. Jamal, Y Kumar, M. Gupta, P. Gupta, I. Sergeev, H. C. Wille & D. Kumar, Hyperfine Interact **242,** 17 (2021); https://doi.org/10.1007/s10751-021-01736-4

[31] H. C. Wille, H Franz, R Röhlsberger, WA Caliebe, FU Dill, J. Phys.: Conf. Ser. **217**, 012008, 2010; https://doi.org/10.1088/1742-6596/217/1/012008

[32] A Gupta, D Kumar, V Phatak, Physical review B **81** (15), 155402, 2010; https://doi.org/10.1103/PhysRevB.81.155402

[33] A Gupta, D Kumar, C Meneghini, Physical Review B **75** (6), 064424, 2007; https://doi.org/10.1103/PhysRevB.75.064424

[34] L. G. Parratt, Phys. Rev. **95**, 359 (1954); https://doi.org/10.1103/PhysRev.95.359

[35] M.T. Johnson, P.J. H Bloemen, F.J.A. den Broeder and J.j. de Vries, Rep. Prog. Phys. **59**, 1409 (1996); https://doi.org/10.1088/0034-4885/59/11/002

[36] S. Bandiera, R. C. Sousa, B. Rodmacq, and B. Dieny , Appl. Phys. Lett. **100**, 142410 (2012); https://doi.org/10.1063/1.3701585

[37] G. Sharma, R. Gupta, D. Kumar and A. Gupta, J. Phys. D: Appl. Phys. 46, 505302 (7pp) (2013), https://doi.org/10.1088/0022-3727/46/50/505302

[38] A. Tiwari, M. K. Tiwari, M. Gupta, H.-C. Wille, and A. Gupta, Phys. Rev. B 99, 205413 (2019), https://doi.org/10.1103/PhysRevB.99.205413

[39] S. Sumi, Y. Kusumoto, Y. Teragaki, K. Torazawa, S. Tsunashima and S. Uchiyama J. Appl. Phys. **73**, 6835 (1993); https://doi.org/10.1063/1.352452

[40] S Gupta, R Sbiaa, M Al Bahri, A Ghosh, S N Piramanayagam, M Ranjbar and J Akerman, J. Phys. D: Appl. Phys. **51,** 465002 (2018); https://doi.org/10.1088/1361-6463/aae1ec

[41] Y. B. Zhang and J. A. Woollam IEEE Trans. Magn. **31**, 3262 (1995); https://doi.org/10.1109/20.490343

[42] H. Kurt, M. Venkatesan and J. M. D Coey, J. Appl. Phys. **108**, 073916 (2010); https://doi.org/10.1063/1.3481452

[43] S. Bandiera, R. C. Sousa, B. Rodmacq, and B. Dieny, Appl. Phys. Lett. **100**, 142410 (2012); https://doi.org/10.1063/1.3701585

[44] R. Röhlsberger, K. Schlage, T. Klein, and O. Leupold, Phys. Rev. Lett. **95,** 097601 (2005); https://doi.org/10.1103/PhysRevLett.95.097601

[45] M. A. Andreeva, Hyperfine Interact **185**, 17–21 (2008); https://doi.org/10.1007/s10751-008-9806-6

[46] http://www.esrf.eu/Instrumentation/software/data-analysis/OurSoftware/REFTIM-1

[47] D. A. Gilbert, J.-W. Liao, B. J. Kirby, M. Winklhofer, C.-H. Lai and K. Liu , Sci. Rep. **6**, 32842 (2016); https://doi.org/10.1038/srep32842

[48] B. Laenens, N. Planckaert, J. Demeter, M. Trekels, C. L'abbé, C. Strohm, R. Rüffer, K. Temst, A. Vantomme, and J. Meersschaut, Phys. Rev. B **82**, 104421 (2010); https://doi.org/10.1103/PhysRevB.82.104421

[49] T. N. Anh Nguyen, Y. Fang, V. Fallahi, N. Benatmane, S. M. Mohseni, R. K. Dumas, and Johan Åkerman, Appl. Phys. Lett. **98**, 172502 (2011); https://doi.org/10.1063/1.3580612

[50] Sebastiaan van Dijken, Jerome Moritz, and J. M. D. Coey, J. Appl. Phys. **97**, 063907 (2005); https://doi.org/10.1063/1.1861964

[51] Y. B. Li , Y. F. Lou , L. R. Zhang, B. Ma, J. M. Bai, F. L Wei, J. Magn. Magn. Mater. **322,** 3789–3791 (2010); http://doi.org/10.1016/j.jmmm.2010.07.047

[52] L. Liu, H. Lv, W. Sheng, Y. Lou, J. Bai, J. Cao, B. Ma, F. Wei, Appl. Surf. Sci. **258**, 5770–5773 (2012); http://doi.org/10.1016/j.apsusc.2012.02.090

[53] Z. Liu, W. Li, W. Fei, and D. Xu, J. Nanomater. **2012**, 174735 (2012); https://doi.org/10.1155/2012/174735